\documentclass[12pt,preprint]{aastex}

\received{}
\accepted{}

\shorttitle{Arecibo HI Pulsar Absorption}
\shortauthors{Weisberg et al}
\usepackage{subfigure}

\begin{document}

\title{Arecibo HI Absorption Measurements of Pulsars and the Electron
Density at Intermediate Longitudes in the First Galactic Quadrant}

\author{J. M. Weisberg\altaffilmark{1}, S. Stanimirovi{\' c}\altaffilmark{2}, 
K. Xilouris\altaffilmark{3}, A. Hedden\altaffilmark{1,3}, A. de la 
Fuente\altaffilmark{1},  S. B. Anderson\altaffilmark{4}, and
F. A. Jenet\altaffilmark{4,5}}
\altaffiltext{1}{Department of Physics and Astronomy, Carleton College,
    Northfield, MN 55057}
\altaffiltext{2}{Department of Astronomy, University of California,
Berkeley, CA 94720}
\altaffiltext{3}{Department of Astronomy, University of Arizona,
Tucson, AZ 85721}
\altaffiltext{4}{Department of Astronomy, California Institute of Technology,
MS 105-24, Pasadena CA 91125}
\altaffiltext{5}{Center for Gravitational Wave Astronomy, University of Texas
at Brownsville, TX 78520}

\slugcomment{Accepted by  ApJ 2007 September}

\begin{abstract}
We have used the Arecibo telescope to measure the HI absorption spectra of eight pulsars.  
We show how  kinematic distance measurements 
depend upon the values of  the galactic constants $R_o$ and $\Theta_o$, and we select our
preferred current values from the literature.  We then
derive kinematic distances for the low-latitude pulsars in our sample and electron densities 
along their lines
of sight.  We combine these measurements with all others in the inner galactic plane visible from 
Arecibo to study the electron density in this region.  The electron density in the interarm range 
$48\degr < l < 70\degr$
 is  $[0.017 \ (-0.007,+0.012)  \ (68\% \ c.l.$)] cm$^{-3}$. This  
is  $0.75\   (-0.22,+0.49)\  (68\% \ c.l.)$ of the value
 calculated by the \citet{cl02} galactic electron density model.  The model agrees more closely with
 electron density measurements toward Arecibo  pulsars lying closer to the galactic center, at $30\degr<l<48\degr$. 
 Our analysis leads to the best current estimate of the distance of
 the relativistic binary pulsar B1913+16:  $d=(9.0\pm3$) kpc.
 
We use the high-latitude pulsars to search for  small-scale structure in the 
interstellar hydrogen observed in absorption over multiple epochs.   PSR B0301+19 exhibited 
significant changes in its absorption spectrum
over 22 yr, indicating HI structure on a $\sim500$ AU scale.

\end{abstract}


\section{Introduction}

Neutral hydrogen (HI) absorption measurements of pulsar signals 
 at $\lambda=21$ cm are important  probes of various
properties of the interstellar medium such as the small-scale structure of of
cold HI \citep{d81,s03b} and calibrators of the
pulsar distance scale and electron density models at large galactic distances \citep{w79,
w87,w95,fw90}. 
These measurements 
are complementary to interferometrically determined
parallaxes, which can be utilized on nearer sources \citep{bET02,cET04}. 
We report new 
Arecibo HI absorption observations of eight pulsars, and we use these
measurements to study the electron density of the interstellar
medium and the small-scale structure of neutral hydrogen clouds.  

Sensitivity limitations indicate that few if any additional pulsar HI
absorption measurements  of electron density in the galactic plane at the intermediate 
first-quadrant galactic longitudes accessible to the 
Arecibo telescope will be made in the next decade or so. Hence it is an appropriate
time to combine the new results with all previous  pulsar HI measurements
 at these longitudes to globally assess the density in the galactic plane in this region.  We will
 use this information to estimate the distance to the relativistic binary pulsar B1913+16.
 In addition, HI absorption measurements on some of our high-latitude pulsars were 
 originally measured over twenty years ago, during which interval the pulsars have moved many
 AU through the interstellar medium. Hence comparison of the old and new absorption
 spectra  yields information on the small scale structure of the absorbing neutral hydrogen.

The plan of the paper is as follows:  The pulsar HI absorption observational technique 
is sketched in the next section (\S2).  We present HI absorption spectra and kinematic distance 
results for 
low-latitude pulsars  in \S3.  In \S4, we use all measured pulsar distances in the inner galactic plane 
visible from Arecibo Observatory to analyze the electron density in that region.  In \S4.1, we review 
the latest 
work on the galactocentric distance of the Sun and the orbital velocity of the Local Standard of Rest
in order to select an optimum model  to use in refining pulsar kinematic distances and galactic
electron densities.   Then
 in \S4.2, we apply the results  of \S4 and \S4.1 to determine the distance of the relativistic binary pulsar  PSR 
B1913+16.  We provide absorption spectra of high-latitude pulsars in \S5, along with analyses of 
time-variability of absorption in those cases where earlier epoch data are available.  Finally,
we discuss our conclusions  in \S6.

\section{Observations}

\setcounter{footnote}{0}

All observations were made with the 305-m Arecibo telescope from 1998
to 2000. The  radio frequency signals near 1420 MHz were mixed to
baseband, sampled, and recorded with  the Caltech Baseband Recorder
\citep{j01}, a 10 MHz  bandwidth, fast-sampling receiver backend.  Every 
100 ns, the CBR sampled 
complex voltages with four-level digitization from the two orthogonally
polarized feed channels, and recorded the samples
to tape for later processing. (For additional details of the observing
techniques and equipment, see  Stanimirovi{\' c} et al
[2003a,b]).  The data were then corrected for quantization \citep{ja98},
Fourier transformed and folded modulo the apparent pulsar period with
the Supercomputers of Caltech's Center for Advanced Computing Research, 
resulting in data cubes consisting of intensity as a function of pulsar
rotational phase (128 phase bins) and radio frequency (2048 frequency 
channels, each having 1 km/s width).  Subsequent processing at Carleton 
College collapsed the data
cubes into two spectra:  the pulsar-{\it{on}} spectrum is a sum
of the
spectra gathered during the pulsar pulse weighted by $I_{psr}(\phi)^2$, while
the sum of those gathered
in the interval between pulses is called the pulsar-{\it{off}} spectrum.  Here
$I_{psr}(\phi)$ is the broadband pulsar intensity in a given pulse phase bin 
$\phi$.

Two final spectra are formed and displayed for each pulsar. The pulsar
{\it{absorption}} spectrum, which represents the spectrum of the pulsar 
alone less any absorption caused by intervening neutral hydrogen, is  the 
normalized difference of the pulsar-{\it{on}} and -{\it{off}} spectra.  In 
order to maximize sensitivity, multiple integrations are summed with a 
weight depending on the square of the pulsar signal strength 
$I_{psr}(t)$ during each integration. In some cases, additional sensitivity
was achieved by  Hanning smoothing the final absorption spectrum, yielding a
resolution of 2 km/s. (Any absorption spectrum that has been Hanning smoothed
is labelled as such when displayed below.)  The HI
{\it{emission}} spectrum is  the  time-integrated,  pulsar-{\it{off}}
spectrum, calibrated in brightness temperature units by matching its peak
 with the Leiden/Argentine/Bonn HI Survey \citep{hb97,aET00,bET05,k05}. All spectra 
 were frequency switched and low-order polynomials 
were fitted to and removed from them in some cases in order to flatten 
intrinsic or scintillation-induced bandpass ripples.

The basic observing parameters are tabulated for low- and high-latitude pulsars in 
the early columns of Tables 1 and 3, respectively.  The quantities 
 $T_{sys,\rm off-line}$ and $\sigma_{\tau,\rm off-line}$,
the system temperature and measured 1-$\sigma$ noise
in optical depth away from the HI line,  are both given.  The former value is the sum of
an estimated 40 K receiver contribution, plus a sky background determined by 
extrapolating the 408 MHz sky temperature \citep{hET82} at the pulsar position 
to 1420 MHz with a spectral index of -2.6.
The HI emission line itself contributes significantly to the system temperature (and hence
to the noise) at those 
velocities where it is present.  To determine  the expected optical depth noise at any 
velocity $v$,  one may use $\sigma_{\tau}(v)=
\sigma_{\tau,\rm off-line}\times [T_{HI}(v)+T_{sys,\rm off-line}]/T_{sys,\rm off-line}$, where
$T_{HI}(v)$ is the measured brightness temperature of HI at that velocity.

\section{Kinematic Distance Analyses and Results on Low-Latitude Pulsars}

A low-latitude pulsar HI absorption spectrum can be used to set limits on 
the pulsar's distance, using the ``kinematic'' technique.  A pulsar is 
farther than an HI cloud that absorbs its 
signal, and closer than one that does {\em{not}}.  The latter ``no 
absorption'' 
criterion is impossible to ascertain in real spectra because of the inevitable
presence of noise which could mask weak absorption.  However it was shown 
by \citet{w79} that emission features with $T_b \geq 35$ K almost always
exhibit significant absorption of radiation from background objects.  
Therefore 
subsequent investigators have assigned an upper distance limit only at
the velocity where one finds both no significant absorption, {\em{and}}
an emission feature of $T_b \geq 35$ K.  We use a flat \citet{fbs89} 
galactic rotation model and the IAU galactic constants of $R_o=8.5$ kpc
and $\Theta_o=220$ km/s \citep{kl86} to convert the  radial velocities to distance. 
The resulting model radial velocities are
shown in a panel under each absorption/emission spectrum pair.
We also add  and subtract velocities of 7 km/s to our nominal velocities,
in order 
to derive estimates of the uncertainties in distance limits due to streaming 
and random gas motions in the Galaxy. These procedures are identical to those 
used in the critical evaluation of all such measurements then extant by
\citet{fw90}, and by  
all subsequent pulsar HI absorption experimenters.  Hence our results are
directly intercomparable with these earlier ones. Our derived distance
limits are discussed below for each pulsar, and are summarized in Table 1.
See \S 4.1 for a discussion of modifications to our derived distances for
different values of the Galactic constants.

\begin{figure*}
\includegraphics[scale=0.60,angle=-90 ]{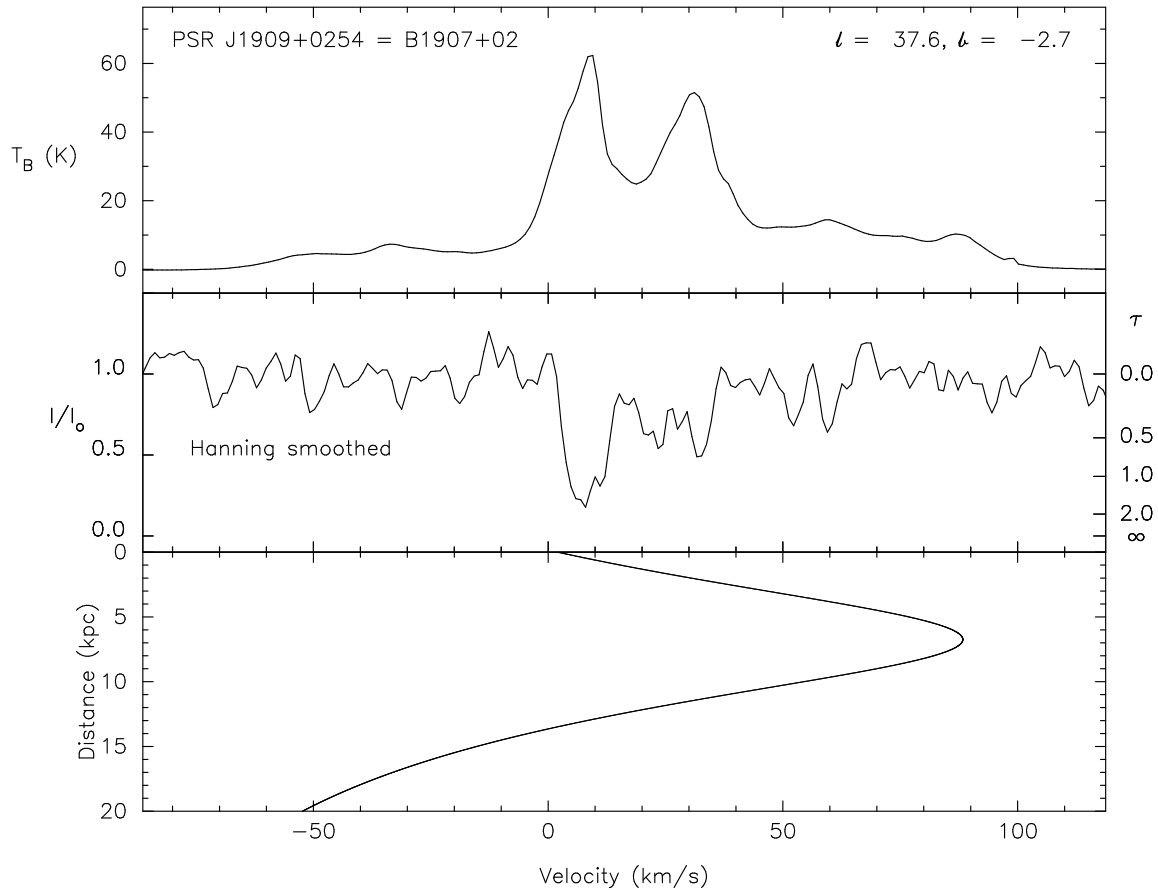}
\caption{  For PSR J1909+0254 = B1907+02.  {\it{Top:}} HI emission spectrum in the 
direction of the pulsar.  {\it{Middle:}} Pulsar
absorption spectrum. The resolution is 1 km/s unless Hanning smoothing is indicated, 
in which
case it is 2 km/s. {\it{Bottom:}} Radial velocity as a function of distance, calculated from a 
\citet{fbs89} flat rotation 
curve with the standard IAU  values of $R_o$ and $\Theta_o$ \citep{kl86}. See 
text for procedures to determine distances with other choices of $R_o$ and $\Theta_o$.}
\label{fig:B1907+02}
\end{figure*}

\subsection{PSR J1909+0254=B1907+02 ({\it{l}}=37\fdg6; {\it{b}}=-2\fdg7), 
Fig. \ref{fig:B1907+02} }

The farthest observed absorption feature is at $ v = 60 $ km/s, well before the
tangent point. Hence the pulsar lies beyond a lower distance limit of
$3.8 \pm 0.5$ kpc.  Unfortunately the HI emission has $T_b \lesssim 10$
K at all velocities where one might test for an upper distance limit via
lack of absorption, because
the $b=2\fdg7$ line of sight rapidly leaves the hydrogen layer.  Since
significant absorption  could  not in any case be guaranteed at the velocity
of such weak emission \citep{w79}, no upper distance limit can be set.

\begin{figure*}
\includegraphics[scale=0.60,angle=-90 ]{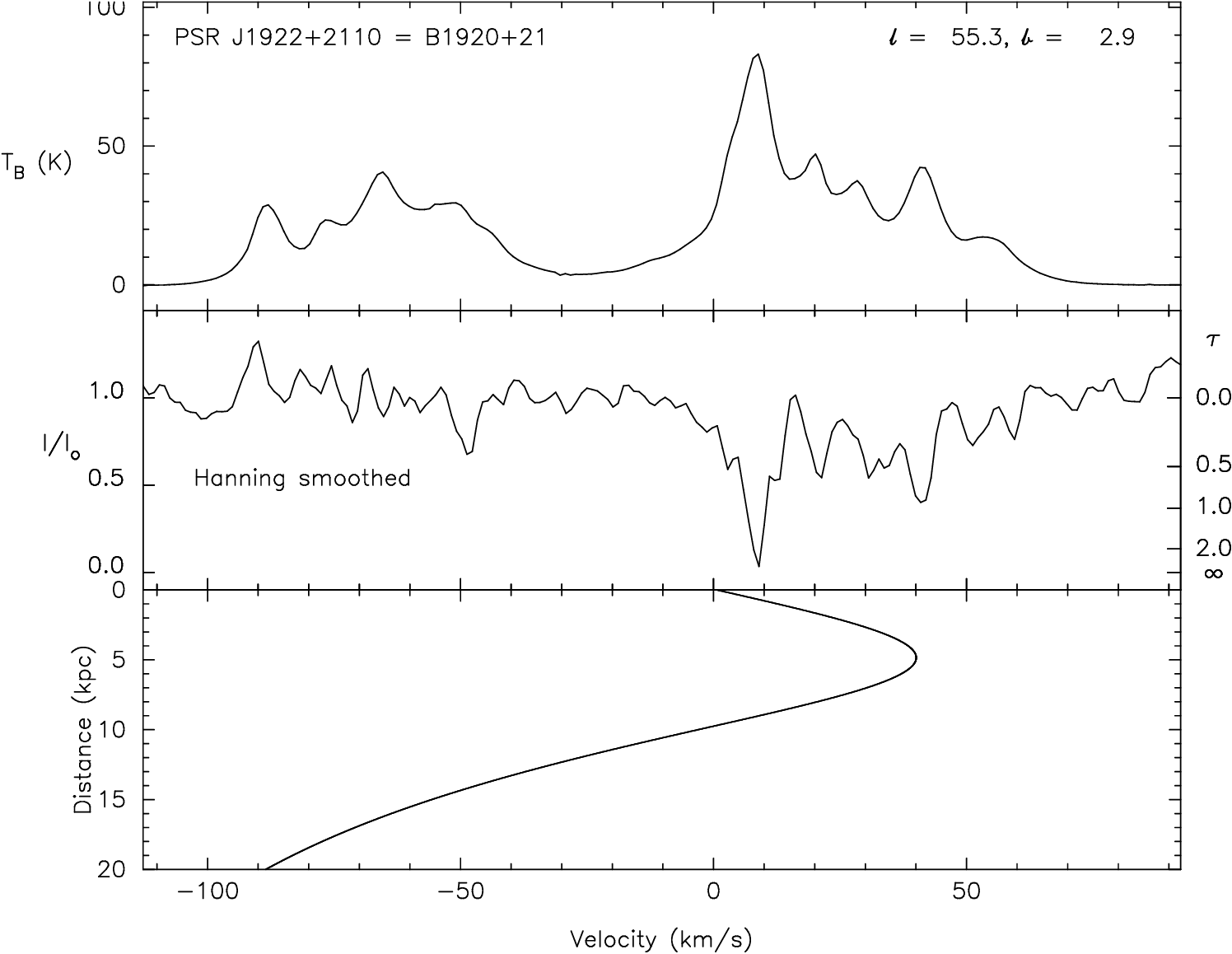}   
\caption{HI emission and pulsar absorption spectra,  and rotation curve, toward  PSR 
J1922+2110 =  B1920+21.  See Fig. 1 caption for 
additional details.}
\label{fig:B1920+21}
\end{figure*}

\subsection{PSR J1922+2110=B1920+21 ({\it{l}}=55\fdg3; {\it{b}}=2\fdg9), 
Fig. \ref{fig:B1920+21} }

\citet{w87} observed this pulsar.  Their HI spectrum was 
contaminated by interstellar scintillation  and was rather
noisy,   prompting
us to reobserve it with greater sensitivity.  The highest velocity
absorption that  \citet{w87}  could confidently detect
was at $v = 26 $ km/s, 
leading to their limit of $(1.9 \pm 0.7$ kpc ) $\lesssim d$.  Our new 
observations clearly exhibit absorption at $v = 41$ km/s, near the
tangent point.   Hence we  revise the lower distance limit significantly
upwards: $(4.8 \pm 1.8$ kpc ) $\lesssim d$.
The \citet{w87} spectrum also showed an absorption feature at this velocity but it 
was not sufficiently above the noise to serve as a reliable kinematic distance
indicator.  The dip in the current absorption spectrum at $v=-48$ km/s is probably 
noise and will not be used to set a distance limit.  (It appears much more
prominently in one of the two circular polarization channels than in the other,
and is not visible in the \citet{w87} absorption spectrum. Furthermore, the nearby
source G55.6+2.3=B1923+210 does not exhibit reliable far side absorption until
$v\lesssim -60$ km/s [Dickey et al 1983; Colgan et al 1988].) Conversely,
the {\it{lack}} of absorption in the $T_b= 41$ K emission feature at $v=-65$ km/s
sets the upper distance limit of $d \lesssim (16.2\pm1.0)$ pc.

\begin{figure*}
\includegraphics[scale=0.60,angle=-90 ]{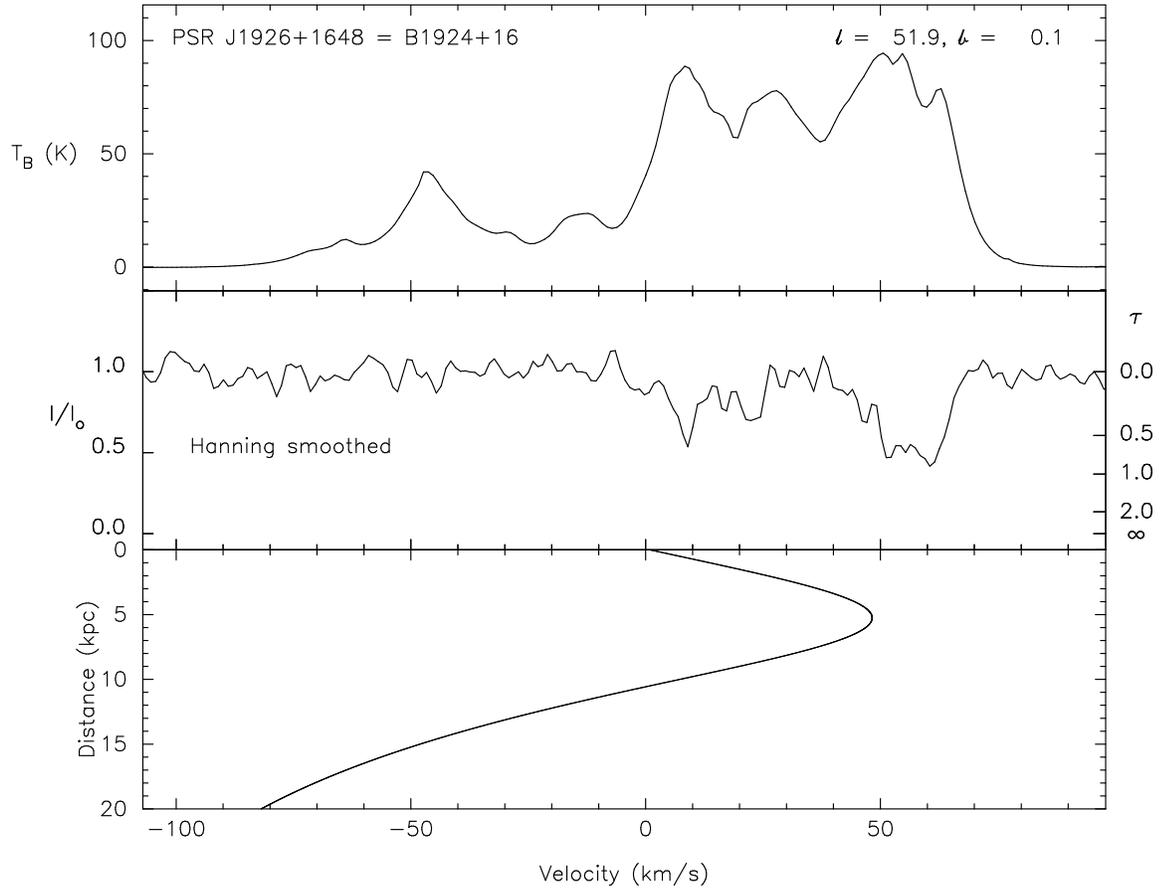}  
\caption{HI emission and pulsar absorption spectra,  and rotation curve, toward  
PSR J1926+1648 = B1924+16.  See Fig. 1 caption 
for additional details.}
\label{fig:B1924+16}
\end{figure*}

\subsection{PSR J1926+1648=B1924+16 ({\it{l}}=51\fdg9; {\it{b}}=0\fdg1), 
Fig. \ref{fig:B1924+16} }

While the Fich et al. model predicts a tangent point velocity of 48 km/s in this
direction, we detect significant emission and absorption well beyond this velocity, 
with 
the last major feature centered near $v=61$ km/s. These features are due to the
Sagittarius arm \citep{b70}.  \citet{gd89} and \citet{cst88} also
observe HI absorption out to these velocities in the nearby sources G50.625-0.031 
and G51.4-0.0, respectively.   We choose the tangent point as our
lower distance limit, so $ (5.2\pm1.8)$ kpc $\lesssim d$. The lack of absorption in
 the $T_b \sim
42$ K emission feature at $v \sim -47$ km/s yields the upper distance limit:  
$ d \lesssim 14.9\pm0.8$ kpc

\section{Analysis of the Electron Density in the Inner Galactic Plane Visible from 
Arecibo}

The currently reported HI absorption distance measurements are probably 
among the last to be determined at Arecibo in the foreseeable future, as all 
known pulsars having sufficient flux density to achieve adequate $S/N$ in a 
reasonable time have now been measured as well as is possible with this 
procedure.   Unfortunately  galactic HI emission is
now the dominant noise source, so that future receiver improvements will not 
significantly decrease the overall noise.  Therefore, new low-latitude pulsars 
discovered in the future in this longitude range will probably be too faint for the HI 
absorption technique, even at Arecibo. Consequently now is a good time to 
review all such measurements and their implications for the electron density
 in the inner galactic plane visible from Arecibo.

The dispersion measure $DM$ of a pulsar, derived from multiple-frequency 
timing measurements and reported in \citet{m05}\footnote{The catalog is 
maintained and updated at 
http://www.atnf.csiro.au/research/pulsar/psrcat}, directly yields the
path-integrated electron density along the line of sight:  $DM = \int{n_e} ds$.
Hence our distance limits, $d$, can  be combined with the published dispersion
measures  to yield mean electron densities $\langle n_e \rangle$ along
the line of sight:

\begin{equation}
\langle n_e \rangle \equiv \frac{ \int{n_e ds} } {d}= \frac {DM} {d}.
\end{equation}

Table 2 lists the electron density limits derived in this fashion from our 
measurements and from  all earlier ones in similar directions; i.e., toward
the inner 
galactic plane accessible to the Arecibo telescope. Note that the kinematic 
distances  listed in Table 2 were all derived 
with  the standard procedures discussed above in \S 3, namely the uniform 
criteria established by \citet{fw90} and the IAU standard 
galactic constants  \citep{kl86}.  Procedures to modify these values for other 
choices of galactic constants are given below in \S4.1.

\begin{figure*}
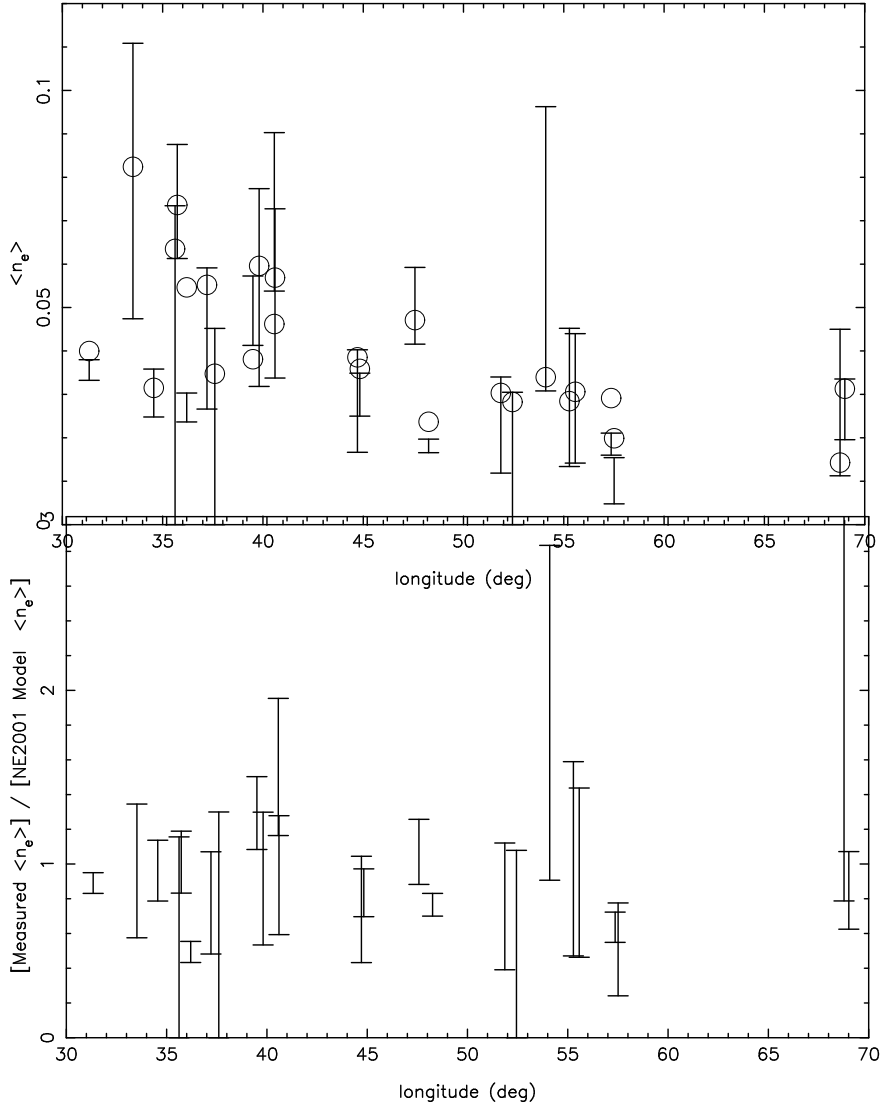

\subfigure{\includegraphics[scale=0.5,angle=-90, trim=1.0in -0.8in 0 0in]{f4a.eps}  }
\subfigure{\includegraphics[scale=0.5,angle=-90, trim=1.0in -0.8in 0 0in]{f4b.eps}  }
\caption{The mean electron density in the inner galactic plane visible from Arecibo 
Observatory. The
IAU \citep{kl86} values of $R_o$ and $\Theta_o$  were used; see text for procedures 
to rescale with
other values.  Pulsars in Table 2 with $d_l <$ 1 kpc or $|b| > 3\degr $ are not shown. 
(a) Measured electron density limits (error bars) and NE2001 model electron densities
(circles), versus longitude.  (b) The ratio of measured electron density to the NE2001 
model value, versus longitude.}
\label{fig:density}
\end{figure*}

\begin{figure*}
\includegraphics[angle=270] {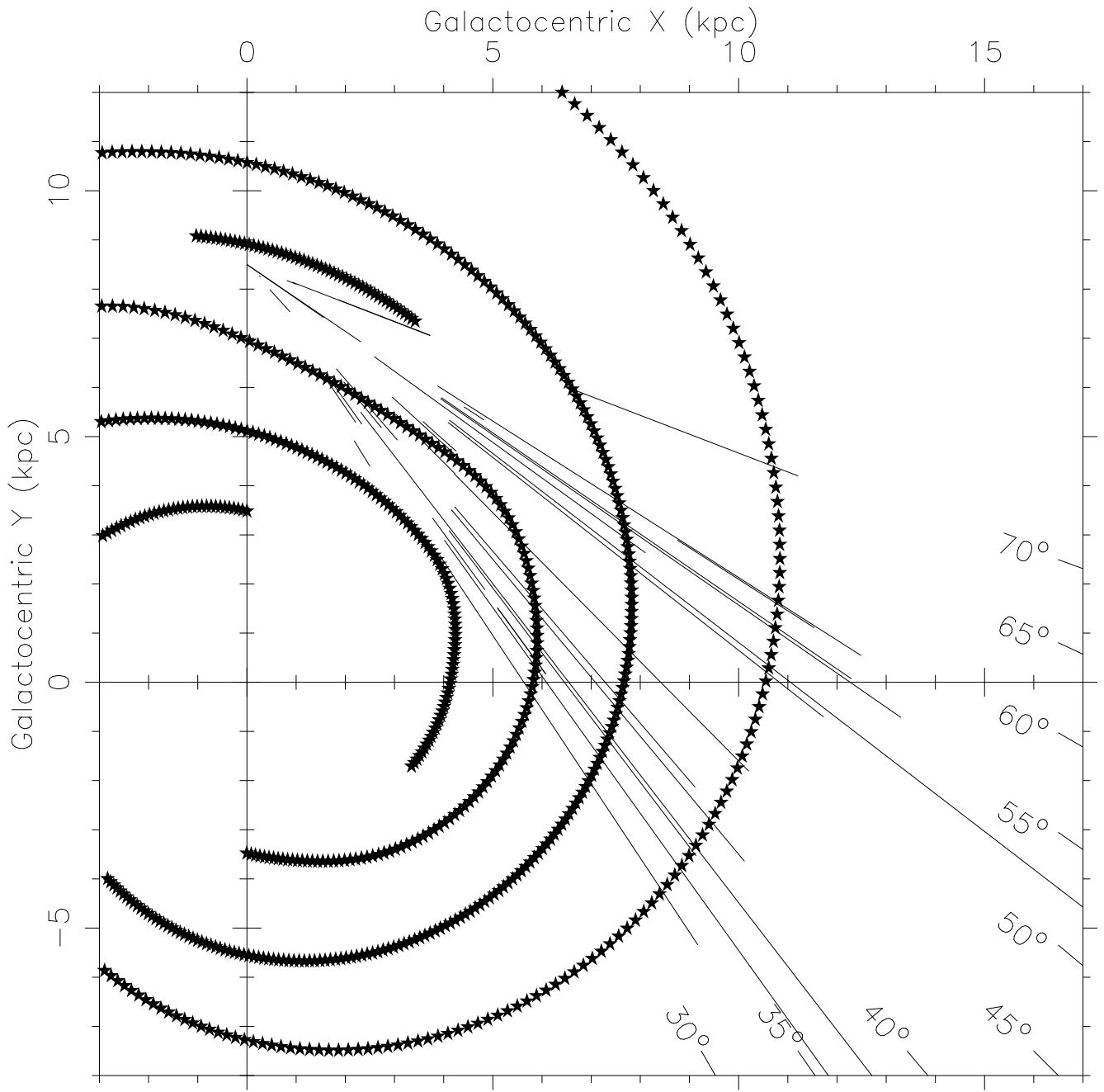}
\caption{The galactic plane, showing spiral arms \citep{cl02} and  Arecibo inner
 galactic plane pulsars with measured distances [upper and lower limits are 
 delineated by the ends of 
lines pointing at the Sun, which is at $(X,Y)=(0.0,8.5)$ kpc].  Galactic longitudes
 in the Arecibo range, $30\degr\le l \le 70\degr$, 
are shown at the bottom and right edges of the plot.   Note especially the long
 interarm path in the direction of PSR B1913+16, at $l\sim50\degr$. }
\label{fig:galplane}
\end{figure*}

In Fig. \ref{fig:density}a, we display measured limits on the mean  electron 
density as a function of galactic longitude  $l$ in the inner galactic plane visible 
from Arecibo.  Only those pulsars from Table 2 with $|b|<3\degr$ and a 
lower distance limit  $d_l \ge 1$ kpc are included in Fig. \ref{fig:density}a
in order to concentrate on kpc-scale averages in the galactic plane.  Almost 
all of these measurements were derived from Arecibo pulsar HI absorption 
spectra.  While there is significant scatter,
it is apparent that the measured densities tend to decline as the line of sight 
rotates away from the inner Galaxy, with a notable drop in the region 
between the Sagittarius arm (the first arm interior to our location) and the local 
spiral arm (see Fig. \ref{fig:galplane}).   \citet{am76} were the first to have 
adequate $\langle n_e \rangle$ 
measurements to infer that interarm densities are lower than those in 
spiral arms.  As illustrated in Fig. \ref{fig:density}a, the numerous measurements 
made since that epoch serve to confirm and refine their suggestion, at least in 
this particular interarm region.  Indeed, the drop in
$\langle n_e \rangle$ as $l$ exceeds $48\degr$ corresponds with the line 
of sight finally reaching a longitude  where it no longer intercepts the 
Sagittarius arm (see Fig. \ref{fig:galplane}). Meanwhile, with  only one 
exception, {\it{every}} pulsar measurement in the interarm range 
$48\degr < l < 70\degr$ is  consistent with a relatively low
 $\langle n_e \rangle\sim0.02$ cm$^{-3}$ (see below for further analysis).  
 The electron density at lower latitudes
sampled  in Fig. \ref{fig:density}a is clearly significantly higher, with an average
in the vicinity of 0.05 cm$^{-3}$ (plus superposed variations)  in the $30-40\degr$ 
longitude range.  

It is also useful to compare our updated set of electron density measurements with
 the most widely used   galactic density model (NE2001, Cordes \& Lazio 2002). 
Fig. \ref{fig:density}b exhibits the measured limits on $\langle n_e \rangle$, 
normalized by the  mean electron densities {\it{predicted}} by the NE2001 model.  
The measured-to-model electron density ratio hovers near
1 (with some inevitable scatter)  for $l \lesssim 48\degr$, indicating that the model 
 fits the data adequately at those longitudes. However, at the higher longitudes
$48\degr < l < 70\degr$ discussed above, there is a tendency for the typical 
measured-to-model density ratio to lie below 1, suggesting that the NE2001 model densities are somewhat high in this interarm region.  

In order to further explore the electron density in this interarm region, we performed
 Monte Carlo simulations to assess the best value and uncertainty on  $\langle n_e \rangle$ 
 and on the measured-to-model $\langle n_e \rangle$ ratio for the pulsars in this region.  
 Of the ten pulsars with measured distance limits and
limits on $\langle n_e \rangle$ in this range (see Table 2), we discarded the data from 
PSR J1935+1616 since it lacks a measured upper distance  and lower density limit.  For each of the
 remaining nine pulsars,  all possible  distances between the measured upper and lower limits
were rendered equally probable in our simulations by choosing distances randomly from a uniform
distribution  between the measured distance limits.  Note that the assumption of uniform probability
is the simplest reasonable hypothesis for measurements such as these which have only a lower
and upper limit.  We find that $\langle n_e \rangle = [0.017 \ (-0.007,+0.012)  \ (68\% \ c.l.$)] cm$^{-3}$, 
and that the
 ratio of measured-to-NE2001 electron density is $0.75\   (-0.22,+0.49)\  (68\% \ c.l.)$ for 
 $48\degr < l < 70\degr$.  Future models should take this discrepancy into account.
 
For simplicity, both the measured and  model densities discussed above are based on the 
standard IAU galactic  constants. As described below in \S4.1,  the {\em{ratio}} 
plotted in Fig.  \ref{fig:density}b remains virtually identical if both constituents are 
rescaled to incorporate other choices of galactic constants.  

\subsection{The Effects of Improved Galactic Constants $R_o$ and $\Theta_o$}

Much progress  has been made on galactic structure and
kinematics in the $\sim 20$ years since the IAU constants were defined. 
Therefore it is useful to discuss improved measurements of $R_o$ and and $\Theta_o$, 
and to investigate the influence  of these 
better values  on the kinematic distances and electron densities discussed in \S3 and \S4.

From remarkable proper motion and radial velocity measurements of the orbit of  star S2  
about Sgr A* through much of its 15-yr period, \citet{eis05}
find that the distance to the galactic center $R_o=7.62\pm0.32$ kpc.  \citet{rb04} have made
equally stunning interferometric measurements of the proper motion of Sgr A*, which yield
an angular velocity of the Local Standard of Rest (LSR) about the galactic center, 
$ \Omega_o = \Theta_o / R_o = 
(29.45 \pm 0.15)$ km s$^{-1}$ kpc$^{-1}$.  Both of these measurements are notable in that 
they rest on far fewer assumptions than earlier determinations, and hence should be freer of 
systematic errors.  Consequently we adopt them in what follows.  An even newer estimate 
of $R_o$ from infrared measurements of red clump 
giants in the galactic bulge \citep{nET06} yields a value of $7.52\pm0.10$ (statistical) 
$\pm 0.35 $ (systematic) kpc, which is consistent with our above adopted value.

 Consider material  traversing a circular  orbit in the plane at galactocentric 
 radius $R$.  Given a flat rotation curve
with circular velocity $\Theta_o$, its radial velocity with respect to the LSR 
is $v_{rad}=\Theta_o(R_o/R-1)\sin l$.  
 With $R=(R_o^2 + d^2 -2 R_o d \cos l)^{1/2}$, the distance  as a function of
  $v_{rad}$  is given by

\begin{equation}
d = R_o  \left( \cos l \pm \sqrt{1+\frac{v_{rad} }{\Theta_o \sin l} }  - \sin^2 l \right).
\end{equation}

Hence our new adopted galactic parameters, giving a small change in $\Theta_o$ but a 
relatively large change in
$R_o$, result in a recalibration of the distance in a particular direction, accomplished  by  
rescaling it with an
 approximately constant  factor of $R_{o,new}/R_{o,old} = 7.62 / 8.50 = 0.896$.  Kinematic 
 distance limits and
 galactic distance models that  use the old IAU galactic constants should be adjusted by 
 this factor to 
 reflect the improved constants.

\begin{figure*}
\subfigure{\includegraphics[scale=0.7]{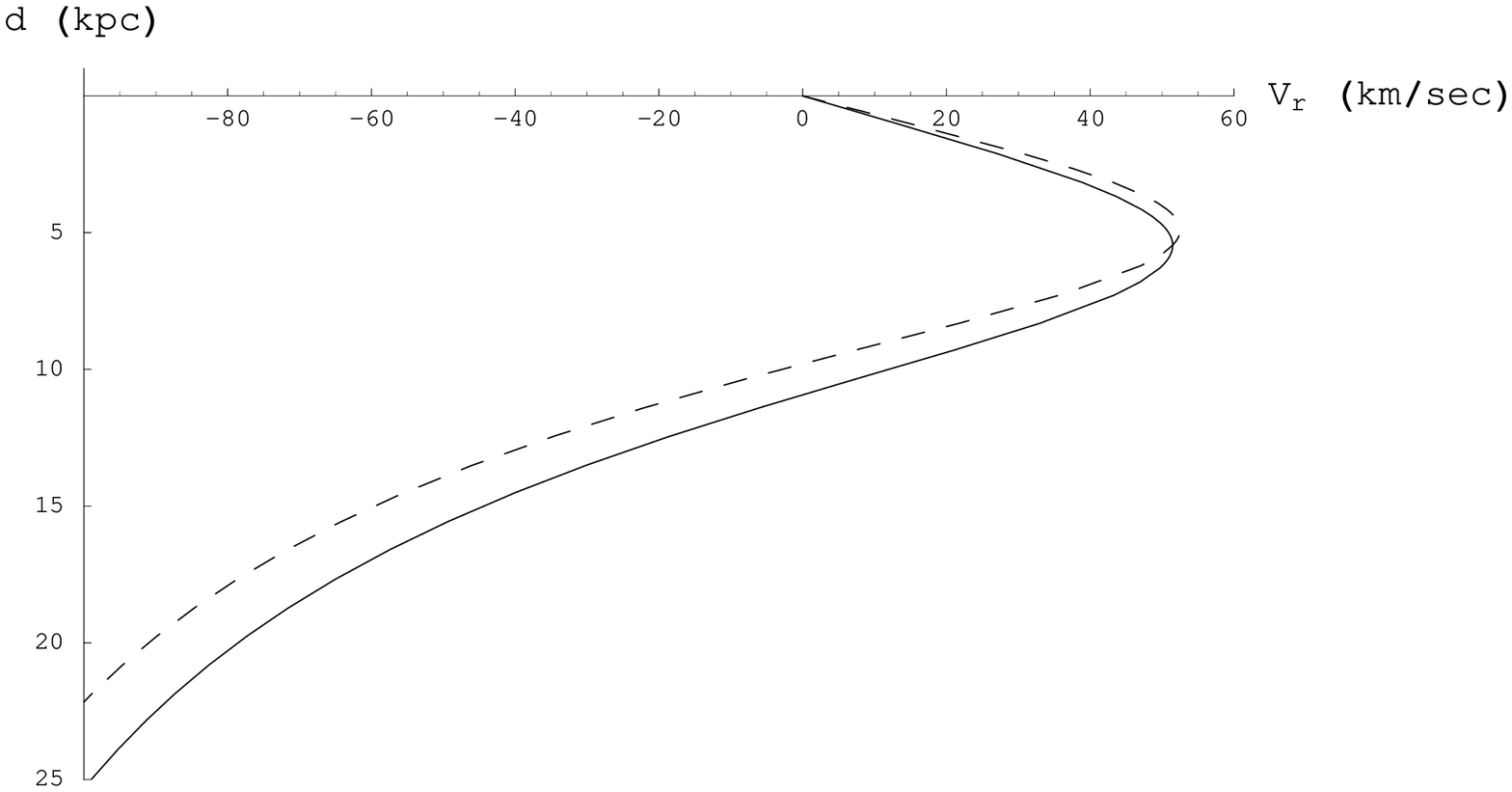}    }
\subfigure{\includegraphics[scale=0.7]{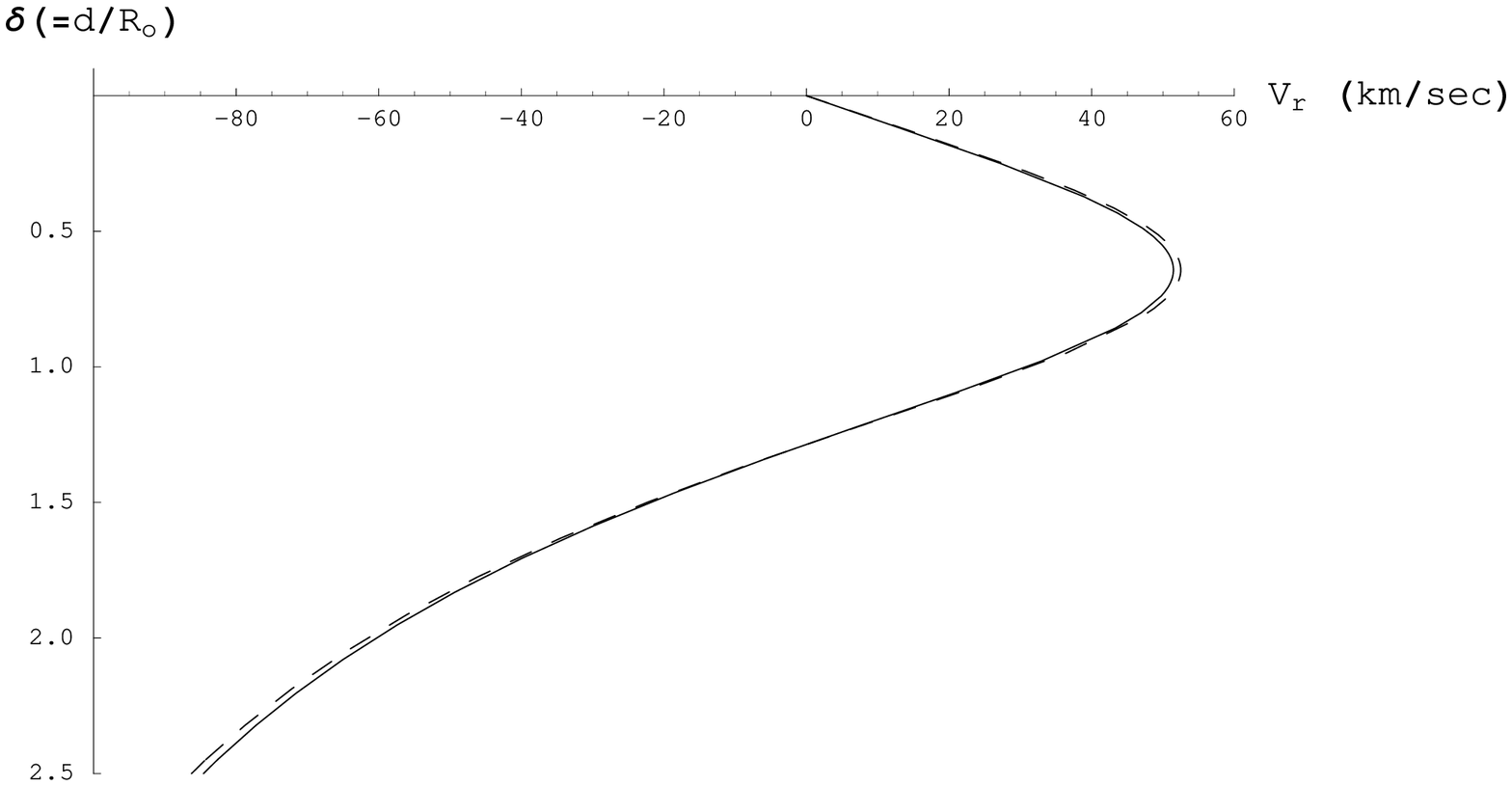}     }
\caption{Calculated radial velocity as a function of distance along a line of sight in
 the galactic plane 
at $l=50\degr$.  A flat rotation curve with the IAU \citep{kl86} values of $R_o$ and 
$\Theta_o$ 
yields the solid line, while a flat rotation curve with new galactic constants (see 
text) is shown by a 
dashed line. (a) Radial velocity versus distance $d$ along the line of
sight.   (b) Radial velocity versus {\it{normalized}} distance $\delta$ along the line of
sight, where $\delta=d/R_o$. }
\label{fig:rotcurve}
\end{figure*}

Fig. \ref{fig:rotcurve} displays  distance -  radial velocity  curves toward $l=50\degr$ derived with 
the old and new flat galactic rotation curves.  Fig. \ref{fig:rotcurve}a illustrates that the 
old and new model
 distances differ primarily by a constant multiplicative  factor, as indicated in the previous 
 analysis.  It follows from the above, that when  distance is {\it{normalized}} by
 $R_o$,  the old and new radial velocity - distance curves are almost identical (Fig. 
 \ref{fig:rotcurve}b).  Such normalized distances are useful in the analysis of the
 acceleration of pulsars in the galactic gravitational field, as will be discussed in the
 next section (\S4.2).

As described above, the new galactic constants would result primarily in the  
measured
distances being rescaled by the ratio of new to old  $R_o$.  The electron 
densities, including
model densities that have been calibrated via pulsar kinematic distance 
measurements,  would
 then be rescaled by the inverse of this ratio (see Eq. 1).  Hence Fig. 
 \ref{fig:density}a  would be thus rescaled; while Fig. \ref{fig:density}b
 would remain unchanged since it displays the ratio of two densities, each 
 of which should be rescaled by the same  factor.

\subsection{The Distance to the Relativistic Binary Pulsar B1913+16}

The relativistic binary pulsar B1913+16 lies in the heart of the region of the 
Galaxy that we are studying, 
at $(l,b)\sim(50\degr,2\degr)$.
The orbital decay of this system due to gravitational wave emission provides 
a strong test of
relativistic gravitation \citep{tw89}.  Currently, the precision of the relativistic 
test is limited by 
uncertainties in the galactic acceleration of the pulsar and the solar system 
\citep{dt91}, which are 
currently dominated by uncertainties in the pulsar distance $d$ [or more specifically,
uncertainties in $\delta$, the distance normalized by $R_o$ \citep{wt03,wt05}].  
Hence it is very important
to assess the current results in an effort to improve our knowledge of B1913+16's 
distance. 

Fig. \ref{fig:density}a shows that  the measured electron density appears to have 
a local minimum near the pulsar's 
longitude.  The relatively low value of $\langle n_e \rangle$  at $l\sim50\degr$ 
is not surprising, since we noted in \S4 that
 the line of sight in this direction traverses a  long path {\it{between}}
spiral arms (see Fig. \ref{fig:galplane}).   The tightest limits on 
 $\langle n_e \rangle$ in the whole
 inner plane visible from Arecibo are for PSR B1915+13 at $l=48\degr$, 
 which happens also to be the
 pulsar closest to PSR B1913+16.  Hence we adopt the measured limits
  from  B1915+13, including
 an allowance for up to  7 km s$^{-1}$ non-circular velocities (not shown in 
 Table 2 or Fig. \ref{fig:density}): 
$\langle n_e \rangle \sim (0.0188 \pm 0.006)\left( \frac{8.5\: {\rm  kpc}}{ R_o}\right) {\rm  cm}^{-3} .$
Note that this value is congruent with and has modestly tighter limits than the electon density 
derived in this interarm region in \S4.

Armed with our new estimate of $\langle n_e \rangle$, we use Equation 1 
to find the distance 
to PSR B1913+16, which has $DM\sim169$ pc cm$^{-3}$:  
\begin{equation}
d_{\rm{B }1913+16}=(10.0\pm3.2) \left( \frac{ R_o}{8.5\: {\rm  kpc}}\right)
\end{equation}
kpc. Our result can also be compared with the \citet{cl02} NE2001 
model distance of  $5.90 \left( \frac{ R_o}{8.5\: {\rm  kpc}}\right) $ kpc, 
which is lower for the reasons discussed above. 
 \citet{dt91} used techniques similar to ours 
to estimate that 
$d_{\rm{B }1913+16}=(9.2\pm1.4)\left( \frac{ R_o}{8.5\: {\rm  kpc}}\right)$.  
While the error bars of their and our estimates
overlap significantly, we are not able to justify tightening them to the degree
that \citet{dt91} did.  Finally,  adopting what we judged in \S4.1 to be the current best 
measurement of $R_o \ (R_o=7.62\pm0.32$ kpc;  Eisenhauer et al. 2005) 
leads to  our best estimate of the distance to PSR B1913+16:  
$d_{\rm{B }1913+16}=(9.0\pm3)$ kpc. 

As noted above, the uncertainty in the general relativistic orbital decay rate of this pulsar is 
currently dominated by the uncertainty in the quantity $\delta_{\rm{B }1913+16}$.  For these 
purposes, it is useful to give 
\begin{equation}
\delta_{\rm{B }1913+16} \equiv (d_{\rm{B }1913+16} / R_o) = (1.18 \pm 0.38).
\end{equation}

\section{High Galactic Latitude Pulsars and Searches for Temporal Variability of Absorption}

Relatively nearby pulsars are
excellent targets for studies of the tiny-scale atomic structure (TSAS, \citet{h97}).
Comparison of their HI absorption spectra at multiple epochs as they move across the
sky permits us to study structure down to AU scales.  We discuss here our
measurements of five such pulsars.  All but two have been
previously observed, so we will search for changes between the previous and
current epochs. The pulsars' observing parameters, distances, and transverse velocities 
are listed in Table 3.

The pulsars discussed in this section  are all located at high galactic latitudes 
(in all cases, $|b| > 25\degr$).  Since their lines of sight leave the galactic hydrogen layer 
fairly quickly, they are not amenable to the kinematic distance technique discussed in \S3.
Conversely, we can not search for temporal absorption variations in the low-latitude pulsars
of \S3 because two have never before been observed, and the third's absorption spectrum
was too noisy at the earlier epoch to yield  meaningful differences.  (Indeed we  reobserved it
in order to more confidently measure its kinematic distance because the noise in the
earlier spectrum precluded a secure result.)

\begin{figure*}
\includegraphics[scale=0.60,angle=-90 ]{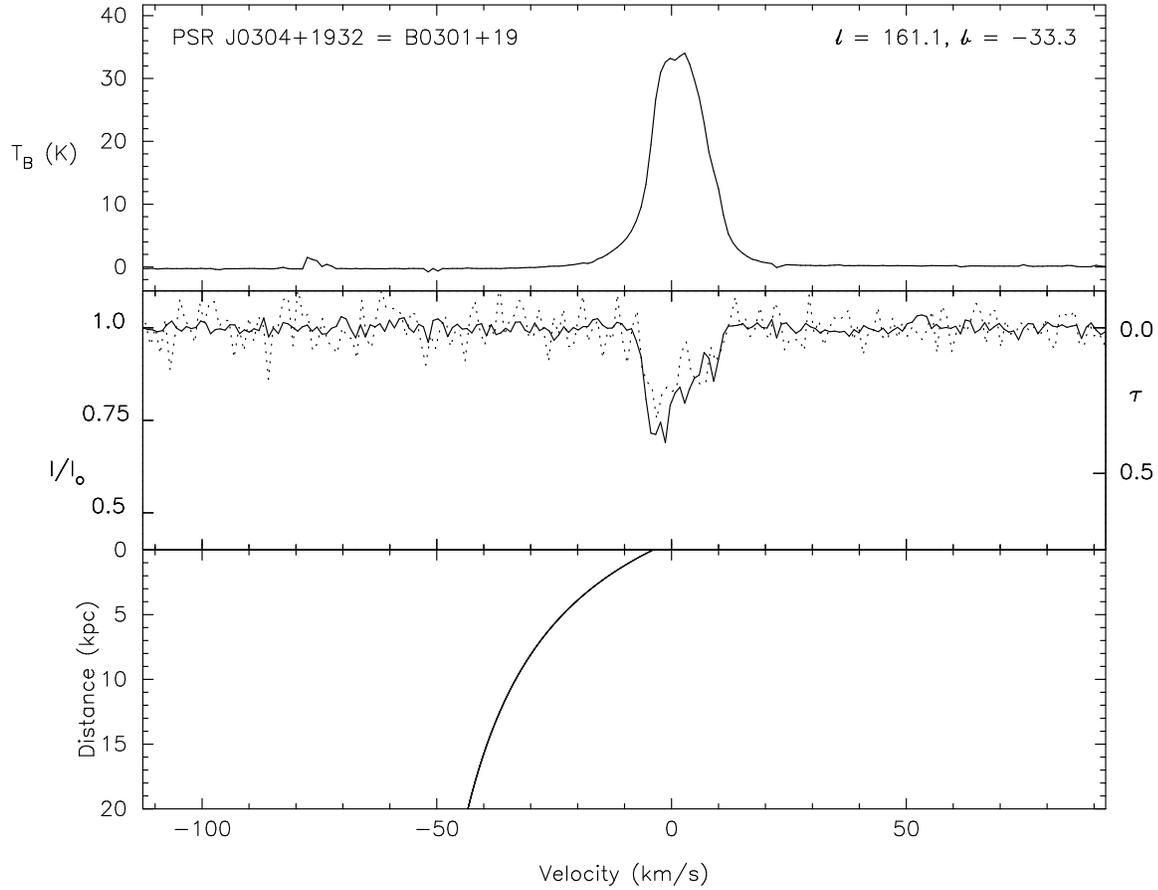} 
\caption{HI emission and pulsar absorption spectra,  and rotation curve, toward  
PSR J0304+1932 = B0301+19.
The 1976-77  absorption spectrum of \citet{d81} is also plotted 
atop ours as a dotted line.  Note that the  absorption spectrum vertical 
scale is magnified  for clarity.  See Fig. 1 caption for additional details.}
\label{fig:B0301+19}
\end{figure*}

\begin{figure*}
\includegraphics[scale=0.70]{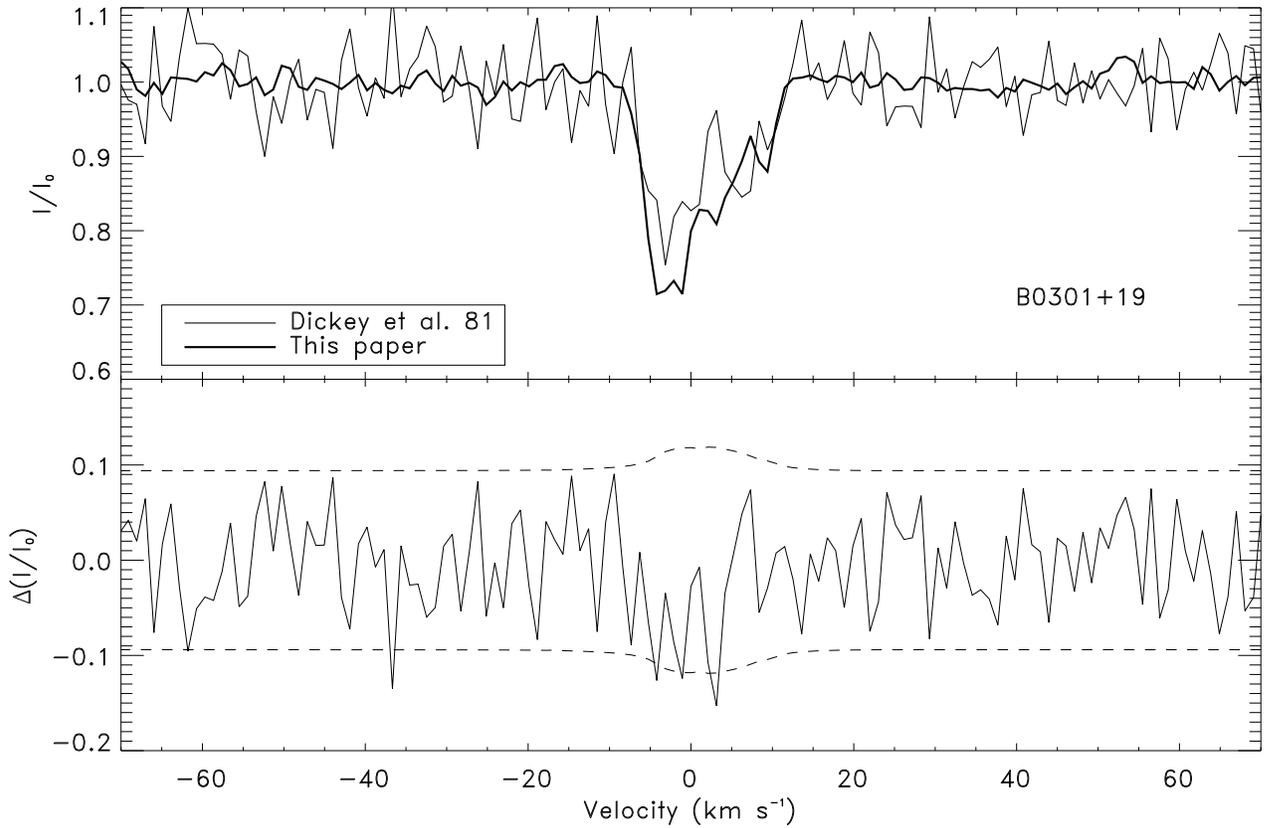}
\caption{Changes in the absorption spectrum of PSR J0304+1932 =  
B0301+19 over twenty-two years. 
(a) The old and new spectra are shown after the new spectrum has been 
interpolated to the same
frequencies as the old spectra.  This plot is similar to the central panel of 
Fig. \ref{fig:B0301+19}, except that the new spectrum 
in the earlier figure was not resampled.  (b) The difference spectrum, with a 
$\pm2\sigma$ error envelope superposed (dashed lines).}
\label{fig:B0301+19diff}
\end{figure*}

\subsection{J0304+1932=B0301+19 ({\it{l}}=161\fdg1; {\it{b}}=-33\fdg3), 
Figs. \ref{fig:B0301+19} and \ref{fig:B0301+19diff}}

The HI absorption spectrum of this pulsar was first observed in 1976-77 
by \citet{d81}. That  spectrum and our new one are plotted together in Fig. 
\ref{fig:B0301+19}.  Both epochs'  spectra have a similar $\sim  1$ km/s  
resolution.  Both the  {\em{integrated}}
 optical depth (the equivalent width)  and the depth of most individual 
absorption components have changed significantly between the two epochs.
  Although the early observations were gathered with a  spectrometer 
that digitized the signal to only one bit, its response to rapidly varying pulsar signals 
was very well-characterized by \citet{w78}.  Careful quantization corrections have also been 
made for the current epoch \citep{ja98}.  Hence we believe that both spectra are accurate 
and that the change is  real. 

The channel spacing of the old and new spectra are   1.056 and 1.031 km/s,  respectively.  In order to 
 study the absorption changes more carefully, we slightly coarsened the spacing of the new data
via interpolation
to match the  channel  spacing and center frequency of the old data. We also ensured that the old and
new channel frequencies matched by  verifying that the (essentially noise-free) {\em{emission}}
 spectra were consistent to less than a channel width.  Fig. \ref{fig:B0301+19diff}  displays the old and 
(resampled) new absorption spectra (top) and their difference (bottom).  The difference spectrum
also displays a $\pm2\sigma$ noise envelope in order to assess the significance of variations.
[The noise envelope grows at the central velocities because the HI emission line itself 
contributes significantly to the system noise \citep{j03,s03b}.]  The difference spectrum clearly
shows a general trend whereby the absorption line depth is greater at the later epoch in  most of the
central channels, with the largest single-channel difference being significant at the 2.6$\sigma$ level. 

The   $\sim22$ year time baseline is unique in being one of
 the longest extant showing absorption variations, leading to a length scale of $\sim500$ AU
 which is second only to the PSR 1557-50 scale of $\sim1000$ AU \citep{j03}. 
Further implications of these results for small-scale 
structure in the interstellar medium will be analyzed in a separate paper 
(Stanimirovic et al, in preparation).

\begin{figure*}
\includegraphics[scale=0.60,angle=-90 ]{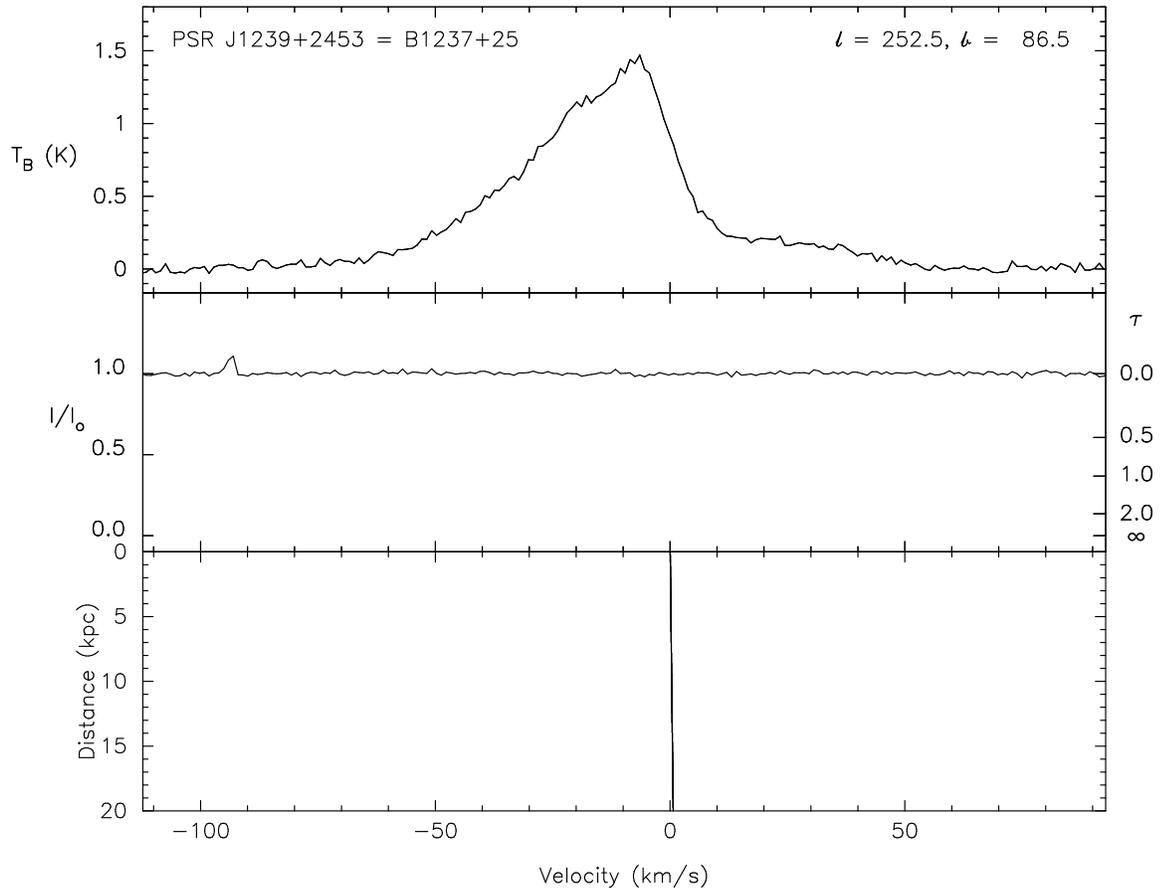}     
\caption{HI emission and pulsar absorption spectra,  and rotation curve, toward  
PSR J1239+2453  = B1237+25. See Fig. 1 caption for details.}
\label{fig:B1237+25}
\end{figure*}

 \subsection{J1239+2453=B1237+25 ({\it{l}}=252\fdg5; {\it{b}}=86\fdg5), Fig. 
 \ref{fig:B1237+25}   }

\citet{d81} also measured the HI absorption spectrum of this pulsar in 1976-77.
The original and current pulsar spectra show no absorption.  These results are
not surprising, given the galactic polar line of sight and the very weak HI emission 
in this direction.  The pulsar moved $\sim2200$ AU between the two observations.

\begin{figure*}
\includegraphics[scale=0.60,angle=-90 ]{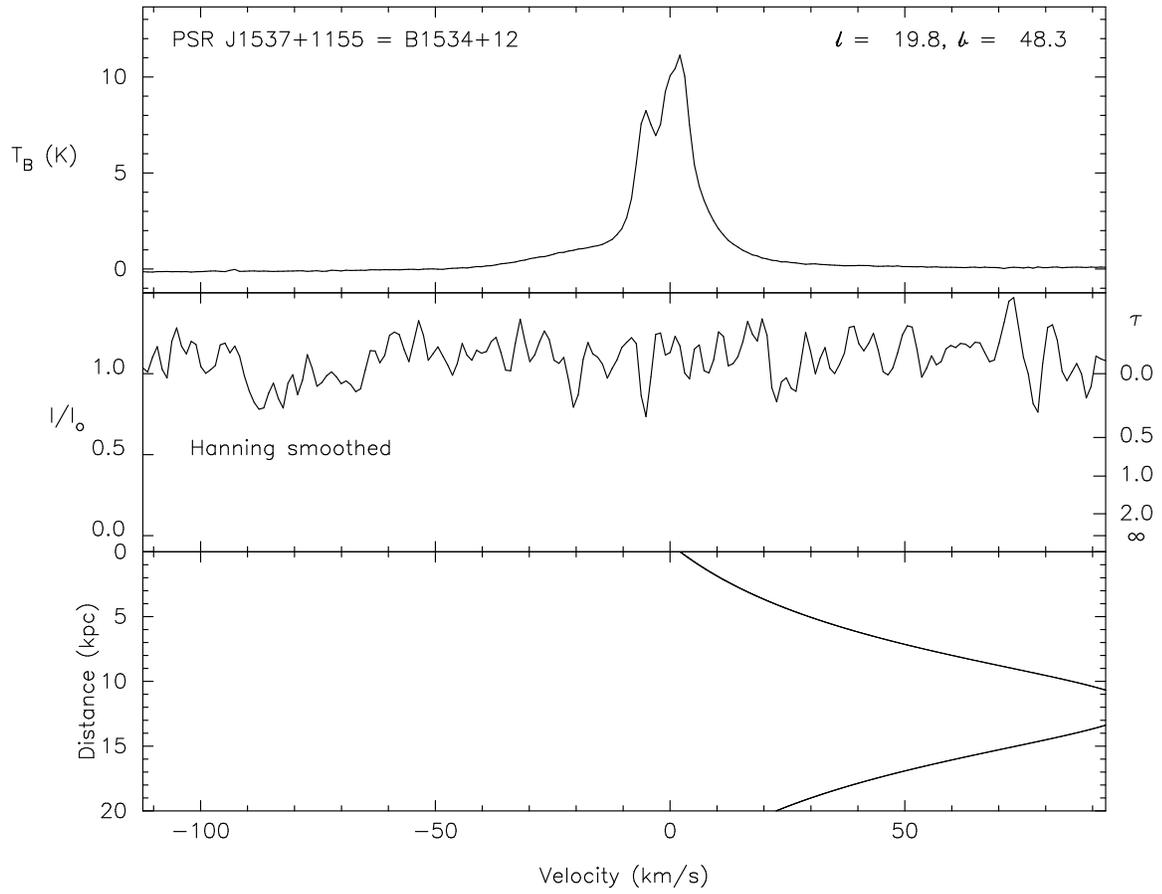}     
\caption{HI emission and pulsar absorption spectra,  and rotation curve, toward  
PSR J1537+1155 =  B1534+12. See Fig. 1 caption for details.}
\label{fig:B1534+12}
\end{figure*}

\subsection{J1537+1155=B1534+12 ({\it{l}}=19\fdg8; {\it{b}}=48\fdg3), Fig. 
 \ref{fig:B1534+12}   }

This pulsar is a member of a double neutron star binary system.  Its pulse
timing parallax gives $d > 0.67$ kpc \citep{s02}.
Under the assumption that general relativity provides the correct description
of gravitational wave emission, the excess observed orbital period change not 
attributable to gravitational wave damping
yields $d=(1.02 \pm 0.05)$ kpc \citep{s02}.  There is no statistically significant
absorption evident in our pulsar spectrum along this  high galactic 
latitude line of sight.  No previous HI absorption measurements have been made
on this source.

\begin{figure*}
\includegraphics[scale=0.60,angle=-90 ]{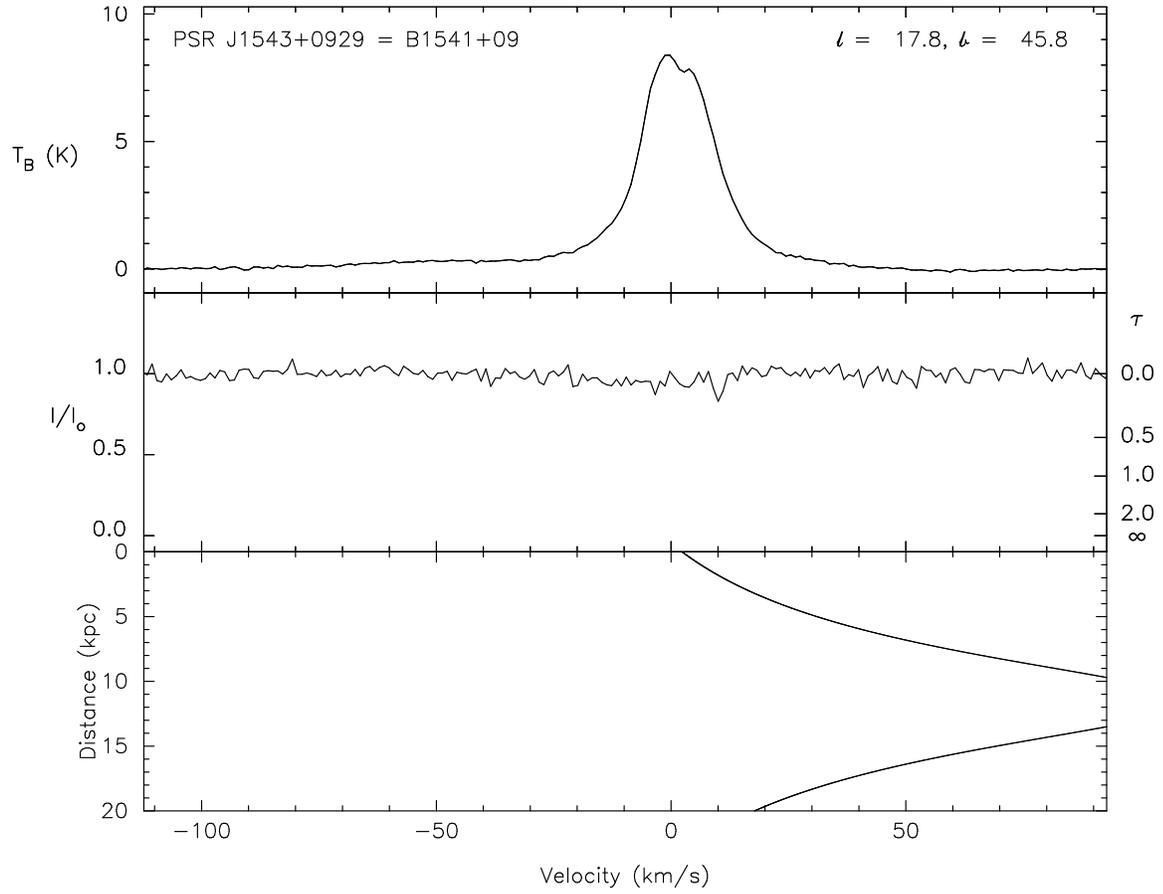}  
\caption{HI emission and pulsar absorption spectra,  and rotation curve, toward  
PSR J1543+0929 =  B1541+09. See Fig. 1 caption for details.}
\label{fig:B1541+09}
\end{figure*}

\subsection{J1543+0929=B1541+09 ({\it{l}}=17\fdg8; {\it{b}}=45\fdg8), Fig. 
\ref{fig:B1541+09}  }

No statistically significant absorption is present in this high galactic latitude
pulsar's spectrum.  This is the first absorption spectrum from this pulsar.

\begin{figure*}
\includegraphics[scale=0.60,angle=-90 ]{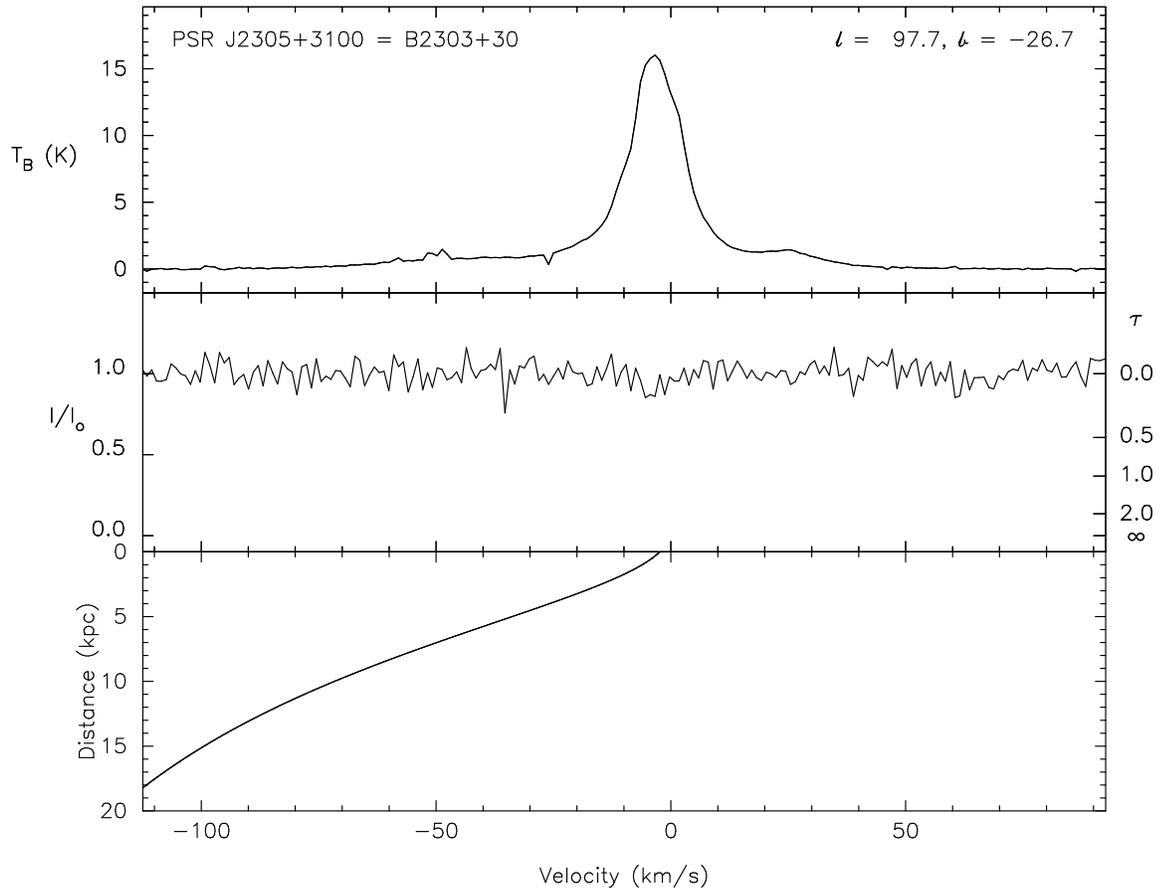}  
\caption{HI emission and pulsar absorption spectra,  and rotation curve, toward  
PSR J2305+3100 =  B2303+30. See Fig. 1 caption for details.}
\label{fig:B2303+30}
\end{figure*}

\subsection{J2305+3100=B2303+30 ({\it{l}}=97\fdg7; {\it{b}}=-26\fdg7), Fig. 
\ref{fig:B2303+30}  }

The broad, shallow dip in the displayed spectrum disappears in 
one of our two orthogonal polarizations, leading us to conclude that it is not real. This 
pulsar was also observed in 1976-77 by \citet{d81}.  In that case as well, no significant
absorption was seen.  The pulsar traversed $\sim1600$ AU in the intervening time.

\section{Conclusions}

We have determined the HI absorption spectra of eight pulsars.  The three low-latitude pulsars
yield kinematic distances and electron densities in the inner galactic plane visible from
Arecibo.  These observations mark the completion of a two-decade effort to accurately 
measure the HI absorption spectrum of all such pulsars that are strong 
enough to be accessible to this technique. Therefore, we have combined our new measurements
 with all others in this direction to study the 
electron density in this region.  We find that the mean electron density in the plane in the interarm 
range $48\degr < l < 70\degr$
 is $ [0.017 \ (-0.007,+0.012)  \ (68\% \ c.l.)$] cm$^{-3}$, which is 
 $0.75\   (-0.22,+0.49)\  (68\% \ c.l.)$ of the \citet{cl02} model value currently used by most researchers.  
 At the lower longitudes  accessible to Arecibo $(30\degr<l<48\degr)$, the \citet{cl02} model appears 
 to conform generally to the measurements, aside from expected local variations.
 
  As part of the process, we show how to modify kinematic distances 
and electron densities as a function of the galactic constants $R_o$ and $\Theta_o$.  We review 
recent efforts to determine the values of these constants and select the best current answers.
Applying  all of these results to the relativistic
 binary pulsar B1913+16, we find a dispersion measure distance of
 $d=(10.0\pm3.2) \left( \frac{ R_o}{8.5\: {\rm  kpc}}\right) $ kpc, or $d=(9.0\pm3$) kpc if we adopt
 our current best choice for $R_o$ \citep{eis05}.

The five high-latitude pulsars are most useful in multiepoch studies of small-scale structure in the
interstellar medium.  Two were observed for the first time by us in this experiment.  Of the
other three, two showed no measurable absorption at either of two epochs separated by 
twenty-two years; while one, PSR B0301+19, exhibited a significant change in absorption profile
over that timespan, indicating HI structure on a $\sim500$ AU scale.

\acknowledgements{JMW, AH, and AF were supported by grant
AST 0406832 from the National Science Foundation.  Arecibo Observatory 
is operated by Cornell University under cooperative agreement with the NSF. }

\newpage

\begin{table}
\caption{ Observing Parameters and Measured Kinematic Distances for Low Latitude Pulsars}
\scriptsize
\begin{flushleft}
\begin{tabular}{cccccccccccc}
\tableline
\tableline
PSR J      &    PSR B   &$t_{obs}$ &$T_{sys, \rm off-line}$&$\sigma_{\tau, \rm off-line}$& {\it{l}} & {\it{b}} &    DM         &$v_l$\tablenotemark{a}
&$v_u$\tablenotemark{a}   & $d_l$\tablenotemark{b}  & $d_u$\tablenotemark{b} \\
           &           &     (hr) & (K) &&(deg)     & (deg)    &(pc cm$^{-3})$ &(km/s)& (km/s) &  (kpc) & (kpc) \\
\tableline
J1909+0254 &  B1907+02 & 3.0 &45&0.11& 37.6  &  -2.7 & 171.7 & 60 & --  & $3.8 \pm 0.5$ & --             \\
J1922+2110 &  B1920+21 & 2.5 &43&0.09& 55.3  &   2.9 & 217.1 & TP\tablenotemark{c} & -65 & $4.8\pm1.8$&$16.2\pm1.0$ \\
J1926+1648 &  B1924+16 & 7.4 &45&0.06& 51.9  &   0.1 & 176.9 & TP\tablenotemark{c} & -47 & $5.2\pm1.8$&$14.9\pm0.8$ \\
\tableline
\end{tabular}
\end{flushleft}
\normalsize
\tablenotetext{a}{$v_l$ and $v_u$ are lower and upper velocity limits, respectively.}
\tablenotetext{b}{$d_l$ and $d_u$ are lower and upper distance limits, respectively. A flat \citet{fbs89} rotation curve 
with the standard IAU  values of $R_o$ and $\Theta_o$ \citep{kl86} was used. }
\tablenotetext{c}{The tangent point velocity is adopted as the lower limit.}

\end{table}

\begin{table}
\caption{  Measured Electron Densities toward Pulsars  with  $30\degr < l < 70\degr; |b| < 3 \degr$; and  1 kpc $\leq d.$ }
\scriptsize
\begin{flushleft}
\begin{tabular}{llcrccccccc}
\tableline
\tableline
PSR J      &    PSR B  & {\it{l}} & {\it{b}} &    DM        
 & $d_l$\tablenotemark{a}  & $d_u$\tablenotemark{a}   &$\langle n_{e,u}\rangle$\tablenotemark{b}&$\langle n_{e,l}\rangle$\tablenotemark{b}   &  Distance&Assoc.  \\
           &           &   (deg)     & (deg)    &(pc cm$^{-3})$ &  (kpc) & (kpc)   &(cm$^{-3})$ & (cm$^{-3})$  & Ref.\tablenotemark{c}  \\
\tableline
J1848-0123    & B1845-01    &  31.34 &   0.04 &  159.5 & 4.2    & 4.8     &  0.038 &  0.033 & FW90   &       \\
J1852+0031   & B1849+00   &  33.52 &   0.02 &  787.0 & 7.1    & 16.6   &  0.111 &  0.047 & FW90    &        \\
J1856+0113   & B1853+01   &  34.56 &  -0.50 &   96.7 & 2.7     & 3.9     &  0.036 &  0.025 & C75  & SNR:W44    \\
J1857+0212   & B1855+02   &  35.62 &  -0.39 &  506.8 & 6.9    &           &  0.073 &              & FW90     &  \\
J1901+0331   & B1859+03   &  37.21 &  -0.64 &  402.1 & 6.8    & 15.1   &  0.059 &  0.027 & FW90    & \\
J1901+0716   & B1859+07   &  40.57 &   1.06 &  252.8 & 2.8    & 4.7     &  0.090 &  0.054 & FW90      &  \\
J1902+0556   & B1900+05   &  39.50 &   0.21 &  177.5 & 3.1    & 4.3     &  0.057 &  0.041 & FW90     & \\
J1902+0615   & B1900+06   &  39.81 &   0.34 &  502.9 & 6.5    & 15.8   &  0.077 &  0.032 & FW90    &  \\
J1903+0135   & B1900+01   &  35.73 &  -1.96 &  245.2 & 2.8    & 4.0     &  0.088 &  0.061 & FW90     &  \\
J1905+0154A  &                     &  36.20 &  -2.20 &  194.0 & 6.4    & 8.2     &  0.030 &  0.024 & R97  & GC:NGC6749 \\
J1906+0641   & B1904+06   &  40.60 &  -0.30 &  472.8 & 6.5    & 14.0   &  0.073 &  0.034 & FW90   &  \\
\textbf{J1909+0254}   & \textbf{B1907+02}   & \textbf{37.60} &  \textbf{-2.71} &  \textbf{171.7} &\textbf{3.8}    &            &  \textbf{0.045} &              & \textbf{this work} &  \\
J1909+1102   & B1907+10   &  44.83 &   0.99 &  150.0 & 4.3    & 6.0    &  0.035 &  0.025 & FW90    &  \\
J1915+1009   & B1913+10   &  44.71 &  -0.65 &  241.7 & 6.0    & 14.5   &  0.040 &  0.017 & FW90   &  \\
J1916+1312   & B1914+13   &  47.58 &   0.45 &  237.0 & 4.0    & 5.7    &  0.059 &  0.042 & FW90     &   \\
J1917+1353   & B1915+13   &  48.26 &   0.62 &   94.5 & 4.8    & 5.7    &  0.020 &  0.017 & FW90    &   \\
\textbf{J1922+2110}   & \textbf{B1920+21}   &  \textbf{55.28} &   \textbf{2.93} &  \textbf{217.1} & \textbf{4.8}    & \textbf{16.2}   &  \textbf{0.045} &  \textbf{0.013} & \textbf{this work}  &  \\
\textbf{J1926+1648}   & \textbf{B1924+16}   &  \textbf{51.86} &   \textbf{0.06} &  \textbf{176.9} & \textbf{5.2}    & \textbf{14.9}   &  \textbf{0.034} &  \textbf{0.012} & \textbf{this work}  &  \\
J1930+1852   &                       &  54.10 &   0.27 &  308.0 & 3.2\tablenotemark{d}    & 10\tablenotemark{d}    &  0.096 &  0.031 & VB88; LAS01  & SNR:G54.1+ 0.3\\
J1932+2020   & B1929+20   &  55.57 &   0.64 &  211.2 & 4.8    & 14.9   &  0.044 &  0.014 & FW90 & \\
J1932+2220   & B1930+22   &  57.36 &   1.55 &  219.2 & 10.4  & 13.7   &  0.021 &  0.016 & FW90   &   \\
J1935+1616   & B1933+16   &  52.44 &  -2.09 &  158.5 & 5.2    &            &  0.030 &             & FW90   & \\
J1939+2134   & B1937+21   &  57.51 &  -0.29 &   71.0 & 4.6     & 14.8   &  0.015 &  0.005 & FW90   &  \\
J1952+3252   & B1951+32   &  68.77 &   2.82 &   45.0 & 1.0     & 4.0    &  0.045 &  0.011   & B84      & SNR:CTB80  \\
J2004+3137   & B2002+31   &  69.01 &   0.02 &  234.8 & 7.0    & 12.0   &  0.034 &  0.020 & FW90    & \\
\tableline
\end{tabular}
\end{flushleft}
\normalsize
\tablenotetext{a}{$d_l$ and $d_u$ are lower and upper distance limits, respectively. A 
flat \citet{fbs89} rotation curve with the standard IAU  values of $R_o$ and $\Theta_o$ \citep{kl86} was used. See 
text for procedures to determine distances with other choices of $R_o$ and $\Theta_o$. }
\tablenotetext{b}{$\langle n_{e,u}\rangle$ and $\langle n_{e,l} \rangle $ are upper and lower mean 
electron density limits, respectively.}
\tablenotetext{c}{Reference key:  B84=\citet{bET84}; C75=\citet{cmr+75}; FW90=\citet{fw90}; 
LAS01=\citet{las01}; R97=\citet{r97}; VB88=\citet{vb88}. Other data are from the ATNF Pulsar 
Catalogue  at http://www.atnf.csiro.au/research/pulsar/psrcat 
\citep{m05}.}
\tablenotetext{d}{Subsequent work by \citet{lET02} suggests $d\sim5$ kpc under an assumption of uniform HI
density in this direction, which the present authors find unlikely. (See Fig. 5.)}

\end{table}

\begin{table}
\caption{ Observing Parameters, Estimated Distances, and Transverse Speeds of High-Latitude Pulsars}
\scriptsize
\begin{flushleft}
\begin{tabular}{ccccccccccc}
\tableline
\tableline
PSR J & PSR B  & $t_{obs}$ &$T_{sys, \rm off-line}$&$\sigma_{\tau, \rm off-line}$& DM &$d$\tablenotemark{a} & $\mu_{tot}$\tablenotemark{b} & Prop. Mo. & $v_{trans}$  &$v_{trans}$ \\
             &              &  (hr)           & (K)&                                                      &(pc cm$^{-3})$ & (kpc)                  & (mas/yr)           &     Ref.\tablenotemark{c}        &(AU/yr)      & (km/s)   \\
\tableline
J0304+1932 & B0301+19 & 1.3 & 41&0.015&15.737    &      0.62          &   37      (5)                            & LAS82    &  23                &   110   \\    
J1239+2453 & B1237+25 &  1.4 &41&0.009& 9.242     &      0.85          &   115.6  (12)                        &   B03       & 98                 &   470   \\ 
J1537+1155 & B1534+12  & 1.7 &42&0.13  & 11.61436 &    1.02         &   25.086 (20)                       &  KWS03 &  26                &   121    \\   
J1543+0929 & B1541+09  & 2.9 &42&0.03& 35.24      &      3.49         &   8.3     (10)                          &   B03      & 29                 & 140    \\  
J2305+3100 & B2303+30  & 1.2 &41&0.07& 49.544   &       3.66         &   20       (2)                           &   B03      & 73                 &  350    \\
\tableline
\end{tabular}
\end{flushleft}
\normalsize
\tablenotetext{a}{Distance $d$ is from the DM and the \citet{cl02} electron density model, except for J1239+2453
[interferometric parallax; \citet{bET02}] and J1537+1155 [differential galactic acceleration (see text); \citet{s02}].}
\tablenotetext{b}{The uncertainty in units of the last digit of the quoted value is given in parentheses.}
\tablenotetext{c}{Reference key:   B03=\citet{b03}; KWS03=\citet{kws03}; LAS82=\citet{las82}.
Other data are from the ATNF Pulsar Catalogue  at http://www.atnf.csiro.au/research/pulsar/psrcat  
\citep{m05}.}

\end{table}

\newpage

\end{document}